\documentclass[a4paper]{jpconf}
\usepackage{graphicx,amsmath,amssymb,amsgen,wasysym,color}
\begin{document}
\title{Chiral Symmetry and Many-Body Effect in Multilayer Graphene}

\author{Yuji Hamamoto}
\address{Institute of Physics, University of Tsukuba, Tsukuba 305-8571, Japan}
\ead{hamamoto.yuji.fp@u.tsukuba.ac.jp}

\author{Tohru Kawarabayashi}
\address{Department of Physics, Toho University, Funabashi 274-8510 Japan}

\author{Hideo Aoki}
\address{Department of Physics, University of Tokyo, Hongo, Tokyo 113-0033, Japan}

\author{Yasuhiro Hatsugai}
\address{Institute of Physics, University of Tsukuba, Tsukuba 305-8571, Japan}
\address{Tsukuba Research Center for Interdisciplinary Materials Science (TIMS), University of Tsukuba, Tsukuba 305-8571, Japan}

\begin{abstract}
Influence of the chiral symmetry on the many-body problem
in multilayer graphene in magnetic fields is investigated.  
For a spinless electron model on the honeycomb lattice 
the many-body ground state is shown to be
a doubly-degenerate chiral condensate 
{\it irrespective of the number of layers}.  
The energy spectrum calculated numerically with the exact diagonalization method 
reveals for ABC-stacked multilayer graphenes that the many-body gap 
decreases monotonically with the number of layers.
\end{abstract}

\section{Introduction}
Graphene's intriguing physics~\cite{RevModPhys.81.109,
RevModPhys.83.407,
RevModPhys.83.1193,
RevModPhys.84.1067} is intimately related to the chiral symmetry,
a fundamental feature of the honeycomb lattice.
The symmetry is defined by the anticommutation relation, $\{\mathcal{H},\Gamma\}=0$, 
with the chiral operator $\Gamma$ acts as
\begin{gather}
\Gamma c_{i}\Gamma=+c_{i}(-c_{i})\quad\mbox{for}\quad i\in\bullet(\circ),\qquad
\Gamma^{2}=1,
\end{gather}
where $\bullet$ and $\circ$ respectively denote the two sublattices.
The symmetry guarantees~\cite{PhysRevB.74.205414,2011JPhCS.334a2004H} the topological stability of the doubled Dirac cones, 
an example of Nielsen-Ninomiya
theorem~\cite{1981NuPhB.185...20N,1981NuPhB.193..173N}, 
as well as the degeneracy of the $n=0$ Landau level (LL), 
which accommodates Aharonov-Casher's argument~\cite{PhysRevA.19.2461}.  
A remarkable phenomenon in the latter is the delta-function-like density of states (DOS) 
for the $n=0$ LL even in the presence of ripples with wavelengths 
exceeding a few lattice constants~\cite{PhysRevLett.103.156804}.
These properties are expected to be inherited
by multilayer graphene~\cite{PhysRevLett.96.086805,PhysRevLett.97.266801,
PhysRevB.78.245416},
since the leading hopping matrix elements, $\gamma_{0},\gamma_{1}$ and $\gamma_{3}$, conserves the bipartite lattice structure, 
and hence the chiral symmetry.  
Indeed, a recent theory~\cite{PhysRevB.85.165410}
predicts that AB-staked bilayer graphene retains 
the anomaly in DOS, which even extends to the case of 
a perpendicular electric field that in fact breaks the chiral symmetry.

Then an interesting question is how the chiral symmetry affects the many-body problem in multilayer graphene.  
Many-body effects have been extensively studied~\cite{RevModPhys.84.1067,
PhysRevLett.96.256602,
PhysRevB.74.075422,
PhysRevB.74.161407,
PhysRevB.74.195429,
PhysRevLett.98.016803,
PhysRevB.75.165411,
%2007SSCom.143..504A,
PhysRevLett.99.196802,
PhysRevB.80.235417,
PhysRevLett.103.216801,
%PhysRevB.81.075427,
%PhysRevB.81.115405,
%PhysRevB.81.205429,
%PhysRevB.85.155439,
PhysRevB.74.161403,
PhysRevB.77.041407,
PhysRevLett.101.097601,
PhysRevB.79.165402,
PhysRevLett.103.266804,
2009EL.....8558005V,
PhysRevB.81.041401,
PhysRevB.81.041402,
PhysRevB.81.075407,
PhysRevB.81.155451,
PhysRevLett.104.156803,
PhysRevB.82.115431,
PhysRevB.82.201408,
PhysRevB.84.235449,
PhysRevB.85.245451,
PhysRevB.77.155416,
PhysRevB.82.035409}
 from the beginning of the graphene research due to interests 
that include the gap opening in the central
LL~\cite{PhysRevLett.96.136806,PhysRevLett.99.106802,PhysRevLett.108.106804,
2012NatPh...8..550Y}
and the unconventional insulating $\nu=0$ state~\cite{PhysRevLett.100.206801}.  
So far, the role of the chiral symmetry in the many-body problem
has been studied in the context of the lattice gauge theory,
where monolayer graphene without magnetic fields is discussed, 
mainly based on the continuum model~\cite{PhysRevLett.102.026802,
PhysRevB.79.165425,
PhysRevB.82.121403}.
In the present work, we consider multilayer graphenes in magnetic fields, 
and investigate the many-body problem in terms of the chiral symmetry.

\begin{figure}
\begin{minipage}{.35\linewidth}
\includegraphics[width=\textwidth]{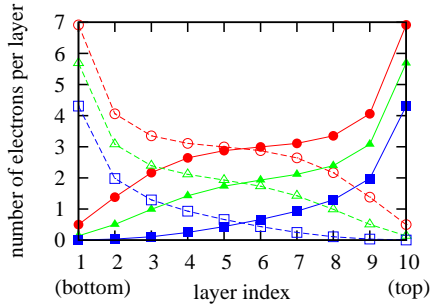}
\end{minipage}
\hfill
\begin{minipage}{.6\linewidth}
\label{fig:charge}
\caption{Layer-by-layer charge distribution
in the chiral condensates for an ABC stack of 10 graphene sheets.
$\gamma_{1}/\gamma_{0}=0.5$,
where only the electrons in the zero-energy LL is shown.
The filled (empty) circles, triangles and squares
correspond to $|G_{+(-)}\rangle$
for $\phi=3/15^{2}, 2/15^{2}$ and $1/15^{2}$, respectively.
}
\end{minipage}
\end{figure}

\section{Ground states}
We consider spinless electrons in multilayer graphenes in a magnetic field,
for which the kinetic energy is 
\begin{gather}
\mathcal{H}_{\rm kin}
=-\sum_{\langle ij\rangle}\gamma_{ij}e^{i\theta_{ij}}c^{\dagger}_{i}c_{j}
+{\rm H.c.},
\end{gather}
where $\gamma_{ij}$ is 
the electron hopping between sites $\langle ij\rangle$.
For the nearest-neighbor in-plane sites $\gamma_{ij}=\gamma_{0}$, 
while for the inter-layer vertical hopping $\gamma_{ij}=\gamma_{1}$.
The magnetic field is included as the Peierls phase $\theta_{ij}$
such that magnetic flux per elementary hexagon equals to 
$\phi=\frac{1}{2\pi}\sum_{\hexagon}\theta_{ij}$ 
in units of the magnetic flux quantum $h/e$.
In the string gauge~\cite{PhysRevLett.83.2246},
the magnetic flux reads $\phi=n/N_{c}$
with an integer $n$ and the number of unit cells, $N_{c}$.
Since the $\gamma_{0}$-$\gamma_{1}$ model preserves 
the chiral symmetry, one can construct a zero-energy multiplet,
$\psi=(\psi_{+},\psi_{-})$ with
$\psi_{\pm}=(\psi_{1\pm},\cdots,\psi_{M_{\pm}\pm})$,
%\begin{gather}
%\psi=(\psi_{+},\psi_{-}),\qquad
%\psi_{\pm}=(\psi_{1\pm},\cdots,\psi_{M_{\pm}\pm}),
%\end{gather}
where $\psi_{m\pm}$'s denote $M_{\pm}$-fold degenerate
eigenstates of $\Gamma$ corresponding to chirality $\pm1$,
i.e., $\Gamma\psi_{m\pm}=\pm\psi_{m\pm}$.
For a $N_{l}$-layer bulk system with a magnetic flux $\phi=n/N_{c}$
$(n=1,2,\cdots)$,
the degeneracy reads $M_{+}=M_{-}=nN_{l}$.
It should be noted that the zero modes with chirality $+$ or $-$
are localized on sublattice $\bullet$ or $\circ$, respectively.
Especially for ABC-stacked multilayers,
the zero modes show sublattiece-selected charge accumulation 
towards the top or bottom layer,
which decays rapidly away from the
surface~\cite{PhysRevB.73.245426,PhysRevB.81.125304,hatsugai-unpub}.
The surface states are analogous to the zigzag edge states
in the ribbon structure~\cite{1996JPSJ...65.1920F}
and should be observed experimentally as local density of states near zero energy
with a scanning tunneling microscope.

\begin{figure}[t]
\begin{minipage}{.3\linewidth}
(a) AB-stacked bilayer\\
\includegraphics[width=\linewidth]{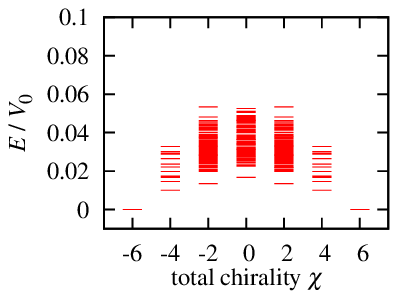}
\end{minipage}
\hfill
\begin{minipage}{.3\linewidth}
(b) ABC-stacked trilayer\\
\includegraphics[width=\linewidth]{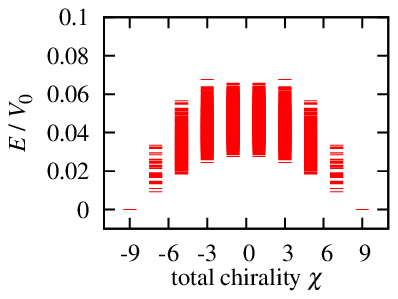}
\end{minipage}
\hfill
\begin{minipage}{.3\linewidth}
(c) ABA-stacked trilayer\\
\includegraphics[width=\linewidth]{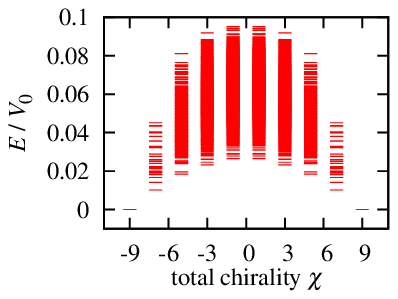}
\end{minipage}
\caption{\label{fig:energy-spectra}
Energy spectra of AB-stacked bilayer graphene (a),
ABC-stacked trilayer graphene (b) and ABA-stacked trilayer graphene (c), 
plotted against the total chirality, $\chi$.  
Results for $N_{c}=15^{2}$, $\phi=3/15^{2}$, $\gamma_{1}/\gamma_{0}=0.5$,
$V_{1}/V_{0}=0.5$ are shown.
}
\end{figure}

For a half-filled multilayer graphene,
the interaction for spinless electrons can be written in a particle-hole symmetric form as
\begin{gather}
\mathcal{H}_{\rm int}
=\sum_{i\ne j}
V_{ij}\left(n_{i}-\frac{1}{2}\right)\left(n_{j}-\frac{1}{2}\right)
=\frac{1}{2}\sum_{i\ne j}V_{ij}
(c^{\dagger}_{i}c^{\dagger}_{j}c_{j}c_{i}
+c_{i}c_{j}c^{\dagger}_{j}c^{\dagger}_{i})
+{\rm const}
\end{gather}
with interaction strength $V_{ij}$.
Since it is hard to treat all the many-body states exactly,
we focus on the many-body problem within the zero-energy LL.
Such a treatment is adequate
as long as
the interaction energy is sufficiently smaller than 
the Landau gap between the central and the adjacent LLs.
The projection onto the zero-energy LL is performed as
$\tilde{c}=(\psi\psi^{\dagger})c$, 
where $c\equiv(c_{1},\cdots, c_{N})^{T}$ with the number of sites $N$,
and $\psi\psi^{\dagger}$ denotes the projection matrix.
Note that the projected creation and annihilation operators are
no longer fermionic; they obey anticommutation relations
$\{\tilde{c}_{i},\tilde{c}_{j}^{\dagger}\}=(\psi\psi^{\dagger})_{ij}$ and
$\{\tilde{c}_{i},\tilde{c}_{j}\}=\{\tilde{c}^{\dagger}_{i},\tilde{c}^{\dagger}_{j}\}=0.$
%\begin{gather}
%\{\tilde{c}_{i},\tilde{c}_{j}^{\dagger}\}=(\psi\psi^{\dagger})_{ij},\qquad
%\{\tilde{c}_{i},\tilde{c}_{j}\}=\{\tilde{c}^{\dagger}_{i},\tilde{c}^{\dagger}_{j}\}=0.
%\end{gather}
In what follows, we assume that the electron-electron 
repulsion only acts between electrons on different sublattices 
[$V_{ij}>0$ for $(i,j) = (\bullet,\circ)$].
The simplest example is the nearest-neighbor repulsion,
which is the leading term for spinless electrons.  
The total Hamiltonian in the projected subspace reads, up to a constant, 
\begin{align}
\tilde{\mathcal{H}}
=\sum_{i\in\bullet, j\in\circ}\frac{V_{ij}}{2}
(\tilde{c}^{\dagger}_{i}\tilde{c}^{\dagger}_{j}\tilde{c}_{j}\tilde{c}_{i}
+\tilde{c}_{i}\tilde{c}_{j}\tilde{c}^{\dagger}_{j}\tilde{c}^{\dagger}_{i})
=\frac{1}{2}\sum_{klmn}(V_{klmn}d^{\dagger}_{k+}d^{\dagger}_{l-}d_{m-}d_{n+}
+V^{\ast}_{klmn}d_{k+}d_{l-}d^{\dagger}_{m-}d^{\dagger}_{n+}),\label{eq:projected-ham}
\end{align}
where $d^{\dagger}_{m\pm}\equiv c^{\dagger}\psi_{m\pm}$
is the creation operator for the zero mode $\psi_{m\pm}$, 
while the pseudopotential appearing for $d, d^{\dagger}$ is defined as
$V_{klmn}=\sum_{i\in\bullet,j\in\circ}V_{ij}
(\psi_{k})^{\ast}_{i}(\psi_{l})^{\ast}_{j}(\psi_{m})_{j}(\psi_{n})_{i}$.
%\begin{gather}
%V_{klmn}=\sum_{i\in\bullet,j\in\circ}V_{ij}
%(\psi_{k})^{\ast}_{i}(\psi_{l})^{\ast}_{j}(\psi_{m})_{j}(\psi_{n})_{i}.
%\end{gather}
Clearly, $\tilde{\mathcal{H}}$ is semi-positive definite;
for an arbitrary many-body state $|\Phi\rangle$, 
$\langle\Phi|\tilde{\mathcal{H}}|\Phi\rangle
=\frac{1}{2}\sum_{i\in\bullet,j\in\circ}V_{ij}
\left|[\tilde{c}_{i}\tilde{c}_{j}
+(\tilde{c}_{i}\tilde{c}_{j})^{\dagger}]|\Phi\rangle\right|^{2}\ge0.$
%\begin{gather}
%\langle\Phi|\tilde{\mathcal{H}}|\Phi\rangle
%=\frac{1}{2}\sum_{i\in\bullet,j\in\circ}V_{ij}
%\left|[\tilde{c}_{i}\tilde{c}_{j}
%+(\tilde{c}_{i}\tilde{c}_{j})^{\dagger}]|\Phi\rangle\right|^{2}\ge0.
%\end{gather}
Since $\tilde{\mathcal{H}}$ commutes with
$\mathcal{G} \equiv \sum_{m}(d^{\dagger}_{m+}d_{m+}-d^{\dagger}_{m-}d_{m-})$,
eigenstates of $\tilde{\mathcal{H}}$ are classified according to the 
total chirality, $\chi=\langle\mathcal{G}\rangle$.
The highest and lowest sectors in $\chi$ correspond to 
{\it chiral condensates}, 
\begin{gather}
|G_{\pm}\rangle=d^{\dagger}_{1\pm}\cdots d^{\dagger}_{M_{\pm}\pm}|D_{<}\rangle,\qquad
\mathcal{G}|G_{\pm}\rangle=\pm M_{\pm}|G_{\pm}\rangle,
\end{gather}
where all the zero modes with chirality $+$ (or $-$) are occupied,
while others are empty.
The chiral condensates are in fact the many-body ground states,
since they are annihilated by both of 
$d^{\dagger}_{k\pm}d^{\dagger}_{l\mp}$ and $d_{m\mp}d_{n\pm}$,
hence $\langle G_{\pm}|\tilde{\mathcal{H}}|G_{\pm}\rangle=0$.
Thus the chiral condensates form a ground-state {\it doublet}, 
$\Psi=(|G_{+}\rangle,|G_{-}\rangle)$,
so that any unitary-mixture, 
$\Psi\mapsto\Psi^{\omega}\omega$ with $\omega\in{\rm U(2)}$, 
is again an eigenstate.  
Since the above discussion holds irrespective of the number of layers,
it is expected that $\Psi$ has properties similar to those 
in the monolayer case, 
such as vanishing Hall conductivity, or the Kekul\'e-type bond-order 
along armchair edges~\cite{hamamoto-unpub1,hamamoto-unpub2}.

As a physical property peculiar to the multilayer case,
we plot in Figure~\ref{fig:charge} layer-by-layer charge distribution
in $|G_{\pm}\rangle$ for an ABC-stacked multilayer
with $N_{c}=15^{2}$ and $N_{l}=10$,
where only the electrons in the zero-energy LL is shown.
Since an experimental cyclotron frequency $\omega_{c}$ is
too small to treat numerically based on our lattice model,
we adopt a rather large parameter $\gamma_{1}/\gamma_{0}=0.5$
to access the realistic parameter region $\gamma_{1}>\omega_{c}$.
One can see charge accumulation towards the top or bottom layer,
which reflects the property of the zero modes
as the surface states~\cite{PhysRevB.73.245426,PhysRevB.81.125304,hatsugai-unpub}.
Notably, a shoulder structure evolves on the intermediate layers with increasing $\phi$,
Although the non-monotonic behavior disagrees with the results of the continuum model,
where low-energy wave functions decays exponentially away from a surface,
the discrepancy seems to be due to the finite size effect of our lattice model.

\section{Excited states}
\begin{figure}
\begin{minipage}{.35\linewidth}
\includegraphics[width=\textwidth]{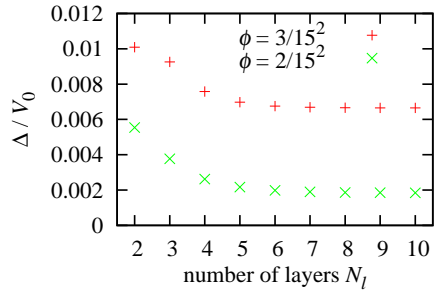}
\end{minipage}
\hfill
\begin{minipage}{.6\linewidth}
\label{fig:gap}
\caption{Energy gap $\Delta$ of ABC-stacked multilayer graphene
composed of $N_{l}$ graphene sheets as a function of $N_{l}$.
Here a result for $\gamma_{1}/\gamma_{0}=0.5, V_{1}/V_{0}=0.5$ is shown.}
\end{minipage}
\end{figure}

Next we investigate excited states at half filling
by numerically diagonalizing
the projected Hamiltonian~(\ref{eq:projected-ham}).
For simplicity, only the nearest-neighbor repulsions
in the same layer $V_{0}$ and those between adjacent layers $V_{1}$
are taken into account.
Fig.~\ref{fig:energy-spectra} shows typical examples of
energy spectra plotted against the total chirality, $\chi$, 
for AB-stacked bilayer graphene and for ABC- or ABA-stacked trilayer graphene.
The spectra are symmetric with respect to $\chi$,
since now we have the degeneracies of the zero modes with chirality $+$ and $-$
are equal.
From this total-chirality-resolved plot 
one can confirm that the ground states are indeed chiral condensates,
as expected from the above discussion.  
More importantly, these data suggest that the first excited states
are obtained by flipping a single-chirality of the chiral condensates as
\begin{gather}
|E_{\pm}\rangle=\sum_{mn}C_{mn}d^{\dagger}_{m\mp}d_{n\pm}|G_{\pm}\rangle
\label{eq:first-excited}
\end{gather}
with coefficients $C_{mn}$ determined numerically.
Although the size of the Hilbert space blows up exponentially
with the number of layers, 
the ansatz~(\ref{eq:first-excited}) enables us to calculate
the energy gap $\Delta$ between the ground state and the first-excited state
by restricting ourselves to the sector of total chirality $\chi=\pm(M_{\pm}-2)$.
Figure \ref{fig:gap} plots thus obtained 
energy gap $\Delta$ of ABC-stacked multilayer graphene
composed of $N_{l}$ graphene sheets.
The decrease in $\Delta$ as a function of $N_{l}$ is consistent with the fact
the overlap of the opposite-chirality zero modes,
which are accumulated near the opposite surfaces,
becomes small with increasing $N_{l}$.
The saturation in $\Delta$ for large $N_{l}$ suggests that
the overlap does not completely vanish due to the finite size effect.

\section{Summary}
We have considered spinless fermions in multilayer graphene
and investigated many-body states in the quantum Hall regime.
For the electron-electron repulsion restricted to unlike sublattices,
the ground state is exactly identified to be chiral condensates
as in the case of monolayer graphene.
Using the exact diagonalization method,
we have numerically calculated the many-body gap,
which decreases monotonically with increasing the number of layers.

\ack
The computation in this work has been done with the facilities of the Supercomputer Center, Institute for Solid State Physics, University of Tokyo. This work was supported in part by Grants-in-Aid for Scientific Research No. 23340112 and No. 23654128 from the JSPS.

\section*{References}

\bibliography{ref}
\end{document}